\documentclass[12pt,showpacs,preprintnumbers,superscriptaddress,amsmath,amssymb,nofootinbib]{revtex4}
\usepackage{graphicx}
\usepackage{dcolumn}
\usepackage{bm}
\usepackage{amssymb}
\usepackage{amsmath}
\usepackage{epsfig}    
\usepackage{color}
\usepackage{hhline}

\def\be{\begin{equation}}
\def\ee{\end{equation}}
\newcommand{\bea}{\begin{eqnarray}}
\newcommand{\eea}{\end{eqnarray}}
\newcommand{\nn}{\nonumber}

\numberwithin{equation}{section}

\begin{document}

\title{Two-loop Neutrino Model with Exotic Leptons}
%
\author{Hiroshi Okada}
\email{macokada3hiroshi@gmail.com}
\affiliation{Physics Division, National Center for Theoretical Sciences, Hsinchu, Taiwan 300}

\author{Yuta Orikasa}
\email{orikasa@kias.re.kr}
\affiliation{School of Physics, KIAS, Seoul 130-722, Korea}
\affiliation{Department of Physics and Astronomy, Seoul National University, Seoul 151-742, Korea}

\date{\today}

\begin{abstract}
We propose a two-loop induced neutrino mass model, in which  we show some bench mark points to satisfy the observed neutrino oscillation, the constraints of lepton flavor violations, and the relic density in the co-annihilation system satisfying the current upper bound on  the spin independent scattering cross section with nuclei. We also discuss new sources of muon anomalous magnetic moment.
\end{abstract}
\maketitle
\newpage

\section{Introduction}
 Nowadays the standard model (SM) becomes trustworthy to describe microscopic fundamental physics, since the SM Higgs has been discovered at the CERN Large Hadron Collider (LHC).
 However it has to be still extended in order to include dark matter (DM) and (minuscule but) massive neutrinos that make an allusion to their existence by overwhelming experimental evidences.
One of the economical and elegant solutions to resolve these controversial issues is to accommodate radiative seesaw models to SM, in which active neutrino masses are radiatively arisen and exotic fields are naturally introduced to induce such tiny masses.
Such an exotic field can be frequently identified as a DM candidate. In this sense, one might find that neutrinos have a strong correlation to the DM candidate.
Following the landmarks~\cite{Zee, Cheng-Li, zee-babu, Krauss:2002px, Ma:2006km, Aoki:2008av, Gustafsson:2012vj},
a vast of literature on radiative seesaw model has recently arisen in Refs.~\cite{Hambye:2006zn, Gu:2007ug, Sahu:2008aw, Gu:2008zf, Babu:2002uu, AristizabalSierra:2006ri, AristizabalSierra:2006gb,
Nebot:2007bc, Bouchand:2012dx, Kajiyama:2013sza,McDonald:2013hsa, Ma:2014cfa, Schmidt:2014zoa, Herrero-Garcia:2014hfa,
Ahriche:2014xra,Long1, Long2, Aoki:2010ib, Kanemura:2011vm, Lindner:2011it, 
Kanemura:2011jj, Aoki:2011he, Kanemura:2011mw,
Schmidt:2012yg, Kanemura:2012rj, Farzan:2012sa, Kumericki:2012bf, Kumericki:2012bh, Ma:2012if, Gil:2012ya, Okada:2012np,
Hehn:2012kz, Baek:2012ub, Dev:2012sg, Kajiyama:2012xg, Kohda:2012sr, Aoki:2013gzs, Kajiyama:2013zla, Kajiyama:2013rla, Kanemura:2013qva, Dasgupta:2013cwa, Baek:2013fsa, Okada:2014vla, Ahriche:2014cda, Ahriche:2014oda,Chen:2014ska,
Kanemura:2014rpa, Okada:2014oda, Fraser:2014yha, Okada:2014qsa, Hatanaka:2014tba, Baek:2015mna, Jin:2015cla,
Culjak:2015qja, Okada:2015nga, Geng:2015sza, Okada:2015bxa, Geng:2015coa, Ahriche:2015wha, Restrepo:2015ura, Kashiwase:2015pra, Nishiwaki:2015iqa, Wang:2015saa, Okada:2015hia, Ahriche:2015loa,
Ahn:2012cg, Ma:2012ez, Kajiyama:2013lja, Hernandez:2013dta, Ma:2014eka, Aoki:2014cja, Ma:2014yka, Ma:2015pma, 
Ma:2013mga, radlepton1, radlepton2, Okada:2014nsa, Brdar:2013iea,
Bonnet:2012kz,Sierra:2014rxa, 
Davoudiasl:2014pya, Lindner:2014oea,Okada:2014nea, MarchRussell:2009aq, King:2014uha}

In this paper we employ new fermions and bosons in addition to the SM-like Higgs boson, in which the leading 
neutrino masses can be induced at the two-loop level where the relevant Lagrangian is controlled by an additional global $U(1)$ symmetry.
And the effective tri-linear coupling between the SM-like Higgs and an isospin triplet boson, which is needed to have a massive CP-odd neutral boson, is also generated at the one-loop level through such exotic fields after the global $U(1)$ symmetry spontaneously.

Following the paper of \cite{Arina:2012aj}, the neutral component can be a DM candidate if there is enough mass difference between two neutral fermions to evade the constraint of the direct detection searches  via the SM neutral gauge boson (Z).
The mass difference arises from the type-II like term after acquiring the vacuum expectation values (VEVs) of the isospin triplet boson.  
However since the mass difference is very tiny because its VEV be less than a few GeV constrained by the electroweak precision test, we have to work on the co-annihilation system to obtain the observed relic density, if we focus on rather lighter DM mass that is less than 80 GeV.

Heavy charged lepton masses themselves are constrained by the LEP  and LHC experiment, and the mass difference between the DM mass and its lepton is also constrained by the electroweak precision test. As a result, the allowed range of the DM mass can be highly restricted.

We have two new sources to explain the deviation of anomalous magnetic moment to SM.
However either of them that has a strong correlation to the neutrino masses cannot reach the sizable value of the anomalous magnetic moment. This is because the neutrino oscillation requires rather large off-diagonal neutrino mass matrix elements, which tends to be in conflict with the anomalous magnetic moment. Notice here that constraints of the lepton flavor violations (LFVs) that always emerge in such radiative models are not so strong to restrict the Yukawa couplings related to the neutrino masses.

This paper is organized as follows.
In Sec.~II, we show our model building including Higgs masses, neutrino mass, LFV, muon anomalous magnetic  moment, and DM.
In Sec.~III, we show our numerical results. We conclude in Sec.~VI.

\section{The Model}


\begin{table}[thbp]
\centering {\fontsize{10}{12}
\begin{tabular}{|c||c|c|c|c|}
\hline Fermion & $L_L$ & $ e_{R} $ & $L'_{L(R)}$  & $e'_{L(R)}$  
  \\\hhline{|=#=|=|=|=|$}
$(SU(2)_L,U(1)_Y)$ & $(\bm{2},-1/2)$ & $(\bm{1},-1)$ & $(\bm{2},-1/2)$   & $(\bm{1},-1)$
\\\hline
$U(1)_{} $ & $-1$ & $-1$ &  $-3/2$ &  $-3/2$    \\\hline
\end{tabular}%
} \caption{Lepton sector; notice the three (or two) flavor index of each field $L_L$, $e_R$, $L'_{L(R)}$ and  $e'_{L(R)}$ is abbreviated.} 
\label{tab:1}
\end{table}

\begin{table}[thbp]
\centering {\fontsize{10}{12}
\begin{tabular}{|c||c|c|c|c|c|}
\hline Boson  & $\Phi$   & $\eta$    & $\varphi$   & $\Delta$    & $S$ 
  \\\hhline{|=#=|=|=|=|=|}
$(SU(2)_L,U(1)_Y)$ & $(\bm{2},1/2)$  & $(\bm{2},1/2)$   & $(\bm{1},0)$   & $(\bm{3},1)$   & $(\bm{1},0)$ \\\hline
$U(1)_{} $ & $0$ & $3/2$ &  $1$ & $3$ &  $1/2$    \\\hline
\end{tabular}%
} 
\caption{Boson sector }
\label{tab:2}
\end{table}

In this section, we review our model, in which
the particle contents for leptons and bosons are respectively shown in Tab.~\ref{tab:1} and Tab.~\ref{tab:2}. 
We add vector-like fermions  of $L'_{L(R)}$ with isospin doublet and $e'_{L(R)}$ with isospin singlet, but $-3/2$ charge under the global symmetry to the SM fields. Each of the exotic field needs (at least) two flavors in order to satisfy current neutrino oscillation data~\cite{pdf}. 
Moreover, an introduction of $e'$ is requested only to have mass difference between the neutral component and the charged component of $L'$ in our model. Otherwise our DM (neutral component of $L'$) does not satisfy the observed relic density $\Omega h^2\approx$0.12~\cite{Ade:2013zuv} due to an enhancement of its thermal averaged cross section with co-annihilation system, as we will discuss later.
As for new bosons, we introduce
two neutral isospin singlet scalars $\varphi$  and $S$  with  $1$ and $1/2$ global charge for each,
an  isospin doublet scalar $\eta$ with  the $3/2$ global charge, and an  isospin triplet scalar $\Delta$ with the $3$ global charge.
Note here that $\Phi$ is neutral under the global charge not to couple to our physical Goldstone boson.
Then we assume that $\Phi$, $\varphi$, and $\Delta$ have  VEV, which are symbolized by $v/\sqrt2$, $v'/\sqrt2$, and  $v_\Delta/\sqrt2$ respectively.
VEV of $\varphi$ spontaneously breaks  the global symmetry down. 
Even after the global $U(1)$ breaking as well as the electroweak breaking, a remnant discrete  symmetry $Z_2$ remains, which is understood as an accidental symmetry. This $Z_2$ symmetry plays a role in  assuring the stability of our  DM candidate; neutral component of $L'_{L(R)}$.

The relevant  Lagrangian for Yukawa sector and scalar potential under these assignments
are given by
\begin{align}
-\mathcal{L}_{Y}
&=
(y_\ell)_a \bar L_{L_a} \Phi e_{R_a} + (y_{L})_{ai} \bar L_{L_a} L'_{R_i} S + (y'_{L})_{ij} \bar L'^c_{L_i}(i\tau_2) \Delta L'_{L_j}
+ (y'_{R})_{ij} \bar L'^c_{R_i}(i\tau_2) \Delta L'_{R_j}\nn\\
&
+(y'_{LR})_{ij}\bar L'_{L_i} \Phi e'_{R_j} 
+(y'_{RL})_{ij}\bar L'_{R_i} \Phi e'_{L_j} 
+ (y_S)_{ib}\bar e'_{L_i} e_{R_b} S^*\nn\\
&
+(M_L)_{ij}  \bar L'_{L_i} L'_{R_j}  +(M_R)_{ij}  \bar e'_{L_i} e'_{R_j} +\rm{h.c.} \label{Lag:Yukawa}\\ 
\mathcal{V}
&=
 m^2_\Phi |\Phi|^2 + m^2_\eta |\eta|^2 + m^2_\varphi |\varphi|^2  + m^2_S |S|^2  + m^2_\Delta {\rm Tr}[|\Delta|^2]\nn\\
&+\mu_S (S^2 \varphi^* +{\rm h.c.}) +\mu_\eta (\eta^T(i\tau_2)\Delta^\dag \eta +{\rm h.c.})
+\lambda_0(\eta^\dag\Phi S\varphi+{\rm h.c.})
 \nn\\
&+
\lambda_{\Phi}|\Phi|^4 + \lambda_{\Phi\eta}|\Phi|^2|\eta|^2  + \lambda'_{\Phi\eta}|\Phi^\dag\eta|^2 + \lambda_{\Phi\varphi}|\Phi|^2|\varphi|^2
+\lambda_{\Phi S}|\Phi|^2|S|^2+\lambda_{\Phi\Delta}|\Phi|^2{\rm Tr}[|\Delta|^2] \nn\\
&+
\lambda'_{\Phi\Delta}\sum_{i}^{1-3}(\Phi^\dag \tau_i\Phi){\rm Tr}[\Delta^\dag \tau_i \Delta]
+
\lambda_{\eta}|\eta|^4 + \lambda_{\eta\varphi}|\eta|^2|\varphi|^2
+\lambda_{\eta S}|\eta|^2|S|^2+\lambda_{\eta\Delta}|\eta|^2{\rm Tr}[|\Delta|^2] \nn\\
&+
\lambda'_{\eta\Delta}\sum_{i}^{1-3}(\eta^\dag \tau_i\eta){\rm Tr}[\Delta^\dag \tau_i \Delta]
+
\lambda_{\varphi}|\varphi|^4 
+\lambda_{\varphi S}|\varphi|^2|S|^2+\lambda_{\varphi\Delta}|\varphi|^2{\rm Tr}[|\Delta|^2] \nn\\
&+
\lambda_{S}|S|^4 
+\lambda_{S\Delta}|S|^2{\rm Tr}[|\Delta|^2] + \lambda_{\Delta}({\rm Tr}[|\Delta|^2])^2 
+\lambda'_\Delta Det[\Delta^\dag \Delta]
,
\label{HP}
\end{align}
where $\tau_i(i$=1-3) is Pauli matrix, each of the index $a(b)$ and $i(j)$ that runs $1$-$3$ and $1$-$2(3)$ represents the number of generations,  and the first term of $\mathcal{L}_{Y}$ can generates the (diagonalized) charged-lepton masses.
We work on the basis where all the coefficients are real and positive for our brevity. 

\subsection{Scalar sector}
After the EW symmetry breaking, each of scalar field has nonzero mass.
We parametrize  these scalar fields as 
\begin{align}
&\Phi =\left[
\begin{array}{c}
\phi^+\\
\phi^0
\end{array}\right],\
\eta =\left[
\begin{array}{c}
\eta^+\\
\eta^0
\end{array}\right],\
\Delta =\left[
\begin{array}{cc}
\frac{\Delta^+}{\sqrt2} & \Delta^{++}\\
\Delta^0 & -\frac{\Delta^+}{\sqrt2}
\end{array}\right].
\label{component}
\end{align}
And the neutral components of the above fields and the singlet scalar field can be expressed as
\begin{align}
\phi^0&=\frac1{\sqrt2}(v + h + ia),\ \eta^0=\frac1{\sqrt2}(\eta_R+i \eta_I),\ 
 \Delta^0=\frac1{\sqrt2}(v_\Delta+\Delta_R+i \Delta_I),\nn\\
 \varphi &= \frac1{\sqrt2}(v'+\rho)e^{i G/v'}, S = \frac1{\sqrt2}(S_R+i S_I),
\label{Eq:neutral}
\end{align}
where $h$ is the SM-like Higgs, and $v$ and $v_\Delta$ is related to the Fermi constant $G_F$ by $v^2+2 v^2_\Delta=1/(\sqrt{2}G_F)\approx(246$ GeV)$^2$.


The CP even Higgs boson mass matrix with VEV in the basis of ($\Delta_R,h,\rho$) is given by
\begin{align}
(M^{2})^{vev}_{\rm CP-even}  = \left(%
\begin{array}{ccc}
\frac{\mu_{\rm eff} v^2}{\sqrt2 v_\Delta} +2\lambda_{\Delta}v_\Delta^2 
&\left[(\lambda_{\Phi\Delta}+\lambda'_{\Phi\Delta})v_\Delta-\sqrt2\mu_{\rm eff}\right]v 
& \lambda_{\Delta\varphi} v'v_\Delta \\
\left[(\lambda_{\Phi\Delta}+\lambda'_{\Phi\Delta})v_\Delta-\sqrt2\mu_{\rm eff}\right]v  
 & 2 \lambda_{\Phi}v^2
 & \lambda_{\Phi\varphi} vv'  \\
 \lambda_{\Delta\varphi} v'v_\Delta & \lambda_{\Phi\varphi} vv'  & 2\lambda_{\varphi} v'^2 \\
\end{array}\right),
\end{align}
and $(M^{2})^{vev}_{\rm CP-even}$ is diagonalized by 3 $\times$ 3 orthogonal mixing matrix $O_R$ as
$O_R (M^{2})^{vev}_{\rm CP-even} O_R^T=$ diag.$(m_{h_1}^2, m_{h_{\rm SM}}^2, m_{h_3}^2)$.
Here
 $h_{\rm SM}$ is the SM Higgs and $h_1$ and $h_3$ are additional Higgses.
eigenstate. 

The CP odd Higgs boson mass matrix with VEV in the basis of ($\Delta_I,a$) is given by
\begin{align}
(M^{2})^{vev}_{\rm CP-odd}  = \left(%
\begin{array}{cc}
\frac{\mu_{\rm eff} v^2 }{\sqrt2 v_\Delta}& -\sqrt2 \mu_{\rm eff} v \\
 -\sqrt2 \mu_{\rm eff} v
&  2\sqrt2 \mu_{\rm eff} v_\Delta \\
\end{array}\right),
\end{align}
and $(M^{2})^{vev}_{\rm CP-odd}$ is diagonalized by 2 $\times$ 2 orthogonal mixing matrix $O_I$ as
$O_I (M^{2})^{vev}_{\rm CP-odd}  O_I^T=$ diag.$(0,m_{a}^2)$, where $m_{a}^2=\frac{\mu_{\rm eff}(v^2+4 v^2_\Delta)}{\sqrt2 v_\Delta}$ and
the massless mode is absorbed by the neutral gauge boson $Z$ to be massive.

The CP even inert Higgs boson mass matrix in the basis of ($\eta_R,S_R$) is given by
\begin{align}
(M^{2})^{inert}_{\rm CP-even}  = \left(%
\begin{array}{cc}
\frac{2m^2_\eta-2\sqrt2\mu_\eta v_\Delta + \lambda_{\eta\varphi}v'^2+(\lambda_{\Phi\eta}+\lambda'_{\Phi\eta})v^2
+(\lambda_{\eta\Delta}+\lambda'_{\eta\Delta})v_\Delta^2 }{2}
&\frac{\lambda_0 v v'}{2}  \\
\frac{\lambda_0 v v'}{2}
& \frac{2m^2_S +2\sqrt2\mu_S v' + \lambda_{\varphi S}v'^2+\lambda_{\Phi S} v^2
+\lambda_{S\Delta} v_\Delta^2 }{2} \\
\end{array}\right),
\end{align}
and $(M^{2})^{inert}_{\rm CP-even} $ is diagonalized by 2 $\times$ 2 orthogonal mixing matrix $V_R$ as
$V_R (M^{2})^{inert}_{\rm CP-even}  V_R^T=$ diag.$(m_{H_1}^2,m_{H_2}^2)$.

The CP odd inert Higgs boson mass matrix in the basis of ($\eta_I,S_I$) is given by
\begin{align}
(M^{2})^{inert}_{\rm CP-odd}  = \left(%
\begin{array}{cc}
\frac{2m^2_\eta + 2\sqrt2\mu_\eta v_\Delta + \lambda_{\eta\varphi}v'^2+(\lambda_{\Phi\eta}+\lambda'_{\Phi\eta})v^2
+(\lambda_{\eta\Delta}+\lambda'_{\eta\Delta})v_\Delta^2 }{2}
&\frac{\lambda_0 v v'}{2}  \\
\frac{\lambda_0 v v'}{2}
& \frac{2m^2_S -2\sqrt2\mu_S v' + \lambda_{\varphi S}v'^2+\lambda_{\Phi S} v^2
+\lambda_{S\Delta} v_\Delta^2 }{2} \\
\end{array}\right),
\end{align}
and $(M^{2})^{inert}_{\rm CP-odd} $ is diagonalized by 2 $\times$ 2 orthogonal mixing matrix $V_R$ as
$V_I (M^{2})^{inert}_{\rm CP-odd}  V_I^T=$ diag.$(m_{A_1}^2,m_{A_2}^2)$.

The singly charged Higgs boson mass matrix with VEV in the basis of ($\Delta^+,\phi^+$) is given by
\begin{align}
(M^{2})^{vev}_{\rm singly}  = \left(%
\begin{array}{cc}
\frac{(\sqrt2\mu_{\rm eff}-\lambda'_{\Phi\Delta} v_\Delta) v^2 }{2 v_\Delta} 
& \left(\frac{\lambda'_{\Phi\Delta}v_\Delta}{\sqrt2}- \mu_{\rm eff}\right) v \\
 \left(\frac{\lambda'_{\Phi\Delta}v_\Delta}{\sqrt2}- \mu_{\rm eff}\right) v
&  (\sqrt2\mu_{\rm eff}-\lambda'_{\Phi\Delta} v_\Delta) v_\Delta \\
\end{array}\right),
\end{align}
and $(M^{2})^{vev}_{\rm singly}$ is diagonalized by 2 $\times$ 2 unitary mixing matrix $O_C$ as
$O_C (M^{2})^{vev}_{\rm singly} O_C^\dag=$ diag.$(0,m_{C}^2)$, where $m_{C}^2=\frac{(\sqrt2\mu_{\rm eff}-\lambda'_{\Phi\Delta}v_\Delta)(v^2+2 v^2_\Delta)}{2 v_\Delta}$ and
the massless mode is absorbed by the charged gauge boson $W^\pm$ to be massive.
The singly charged inert  boson mass eigenstate is given by
\begin{align}
m^2_{\eta^\pm}=\frac{2m^2_\eta+\lambda_{\eta\varphi}v'^2+\lambda_{\Phi\eta}v^2 +(\lambda_{\eta\Delta}-\lambda'_{\eta\Delta}) v^2_\Delta}{2}.
\end{align}

The doubly charged boson mass eigenstate is given by
\begin{align}
m^2_{\Delta^{\pm\pm}}=\left(\frac{\mu_{\rm eff}}{\sqrt2 v_\Delta}-\lambda'_{\Phi\Delta}\right)v^2.
\end{align}

\subsection{Effective trilinear coupling of  $\mu_{\rm eff}$}
In our model, the term $\mu_{\rm eff}\Phi^T(i\tau_2)\Delta^\dag\Phi$ is forbidden at the leading order,
 but induced at the one-loop level mediated by inert neutral bosons $\eta^0$ and $S$ as  depicted in the lower part of Fig.~\ref{neut-mass}.
 The effective trilinear coupling of  $\mu_{\rm eff}$ is given by
 \begin{align}
\mu_{\rm eff}&=\frac{(\lambda_0 v')^2\mu_\eta}{6}\int_0^1\frac{dx_1dx_2dx_3\delta(x_1+x_2+x_3-1)}{(4\pi)^2}\\
&
\left[
\frac{(V_R^T)_{2i}(V_R)_{i1} (V_R^T)_{1j}(V_R)_{j2} (V_R^T)_{1k}(V_R)_{k1}}{x_1 m^2_{H_i}+x_2 m^2_{H_j}+x_3 m^2_{H_k}}
-
\frac{(V_I^T)_{2i}(V_I)_{i1} (V_I^T)_{1j}(V_I)_{j2} (V_I^T)_{1k}(V_I)_{k1}}{x_1 m^2_{A_i}+x_2 m^2_{A_j}+x_3 m^2_{A_k}}
\right],\nn
 \end{align}
 where each of ($i,j,k$) runs form 1 to 2.

\subsection{Inert conditions}
To forbid VEVs  for our inert bosons $\eta^0$ and $S$, the quartic couplings of $\lambda_{\eta}$ and $\lambda_S$ has to be always positive.
To achieve the situation up to one-loop level, we have to satisfy the following conditions at least:
\begin{align}
0\le \lambda_{\eta}^{\rm total}=\lambda_\eta+\delta\lambda_\eta^{(1)},\
0\le \lambda_{S}^{\rm total}=\lambda_S+\delta\lambda_S^{(1)},
\label{eq:cond-inert}
\end{align}
where
\begin{eqnarray}
\delta\lambda_\eta^{(1)}=-10\left| \mu_\eta\right|^4 F_0\left(\eta,\Delta\right),
\end{eqnarray}
\begin{eqnarray}
\delta\lambda_S^{(1)}= -8\left| \mu_S\right|^4 F_0\left(S,\varphi\right),
\end{eqnarray}
with
\begin{align}
F_0(f_1,f_2)=\frac{1}{(4\pi)^2}\int_0^1dxdy\delta(x+y-1)\frac{xy}{(x m^2_{f_1} + y m^2_{f_2})^2}.
\end{align}

\subsection{Fermion Sector}
Let us fist define the exotic  fermion as follow:
\begin{align}
L'_{L(R)}\equiv 
\left[
\begin{array}{c}
N'\\
E'^-
\end{array}\right]_{L(R)}.
\end{align}
{\it Neutral exotic fermion}:
Then the mass matrix for the neutral fermion in the basis of $[N'_L,N'^c_R]$ is given by 
\begin{align}
M_N=
\left[\begin{array}{cc}
M_L & m \\
m &  M_L \\
\end{array}\right],
\end{align}
where $m\equiv y'_L v_\Delta/\sqrt2$ ($m<<M_L$) and we assume to be one generation case with positive real for our simple analysis.
$M_N$ is diagonalized by 2 $\times$ 2 unitary mixing matrix $V_N$ as
$V_N M_N V_N^T=$ diag.$(M_L-m, M_L + m)$, where $V_N$ is written as a maximal mixing form
\begin{align}
V_N=\frac{1}{\sqrt2}
\left[\begin{array}{cc}
i & -i \\
1 &1 \\
\end{array}\right],
\end{align}
and we define the mass eigenstate of the neutral fermion [$N_1,N^c_2$] then we have the following relations:
\begin{align}
\left[\begin{array}{c}N'_L\\ N'^c_R\end{array}\right]
=
\frac{1}{\sqrt2}
\left[\begin{array}{c}
i  N_1+N^c_2 \\
-iN_1 +N^c_2 \\
\end{array}\right].
\end{align}
Furthermore we redefine these fields as 
$\psi_1\equiv N_1+N^c_1$ and $\psi_2\equiv N^c_2+N_2$, then we obtain the Majorana fields $\psi_1$ and $\psi_2$.
We summarize the relations between $\psi$ and N below
\begin{align}
P_L \psi_1&=N_1,\ P_R \psi_1=N^c_1,\ P_L\psi_2=N^c_2,\ P_R \psi_2=N_2,\\
\bar \psi_1 P_R &=\bar N_1,\ \bar \psi_1 P_L =\bar N^c_1,\ \bar \psi_2 P_R =\bar N^c_2,\ \bar \psi_2 P_L =\bar N_2.
\end{align}
The lighter field $N_1$ can be a DM candidate, but the mass difference between them $2m$ is expected to be tiny because $m$ is originated from $v_\Delta$. As a result, we have to consider the co-annihilation system  (at least) including $\psi_1$ and $\psi_2$ to obtain the relic density.
Notice here that the lowest bound on $m$ comes from the inelastic scattering through $Z$ boson and should be heavier than ${\cal O}$(100) keV~\cite{Arina:2012aj}.

{\it Singly charged exotic fermion}:
The mass matrix for the singly charged fermion in the basis of $[E', e']$ is given by 
\begin{align}
M_E=
\left[\begin{array}{cc}
M_L & m'_e \\
m'_e &  M_R \\
\end{array}\right],
\end{align}
where we assume $m'_e \equiv y'_{LR} v/\sqrt2\approx y'_{RL} v/\sqrt2$ ($m'_e<<M_L M_R$) with one generation. Note here that the mass matrix can be taken positive real without loss of generality.
$M_E$ is diagonalized by 2 $\times$ 2 unitary mixing matrix $V_C$ as
\begin{align}
V_C M_E V_C^T= \frac12{\rm diag.}\left(M_L + M_R-\sqrt{(M_L-M_R)^2+4 m'_e}, M_L + M_R+\sqrt{(M_L-M_R)^2+4 m'_e}\right),
\end{align}
where $V_C$ is as
\begin{align}
V_C=
\left[\begin{array}{cc}
c_E & s_E \\
-s_E &c_E \\
\end{array}\right],\
s_{2E}
=\frac{m'_e}{\sqrt{(M_L-M_R)^2+4 m'_e}},
\end{align}
and $s_E\equiv \sin\theta_E$ and $c_E\equiv \cos\theta_E$.
Then we define the mass eigenstate of the charged fermion [$E_1,E_2$], we have the following relations:
\begin{align}
\left[\begin{array}{c}E'\\ e'\end{array}\right]
=
\left[\begin{array}{c}
c_E E_1- s_E E_2  \\
s_E E_1+c_E E_2 \\
\end{array}\right].
\end{align}

{\it Relation between neutral and charged fermion}:
Since the DM mass ($M_X$) can be approximately given by $M_L$ and $m'_e<<M_L,M_R$,
we  have the relation from the charged fermions approximately:
\begin{align}
M_{E_1}\approx  M_X,\ M_{E_2}\approx  M_R.
\end{align}
On the other hand if the mass difference between $X$ and $E_1$ is enough tiny,
$E_1$ also participates in the co-annihilation system of DM. 
The upper bound on $M_E$ is derived from the electroweak precision data, which is typically written in term of $\Delta S,\Delta T,\Delta U$ parameters. The most stringent bound comes from $\Delta T$ that suggests the mass difference between $M_X$ and $M_E$ should be smaller than 45 GeV~\cite{Arina:2012aj}:
\begin{align}
M_{E_1}-M_X\lesssim 45\ {\rm GeV}.
\end{align}
The mass of $E_1$ is also constrained by the LHC searches and the lower bound is around 300 GeV if $E_1$ decays into the SM charged-lepton + missing(=DM), which is an analogous analysis of the slepton searches~\cite{Okada:2014qsa}.
However $E_1$ cannot decay into the SM charged-lepton + missing kinematically but decay into the SM charged lepton + two missing fermions( active neutrinos  + DM) through the lighter mass of $H_1$ or $A_1$, which is assumed to be heavier than the mass of $E_1$ in our case. Then only the constraint comes from the LEP experiment that suggests~\cite{j.beringer}
\begin{align}
100\ {\rm GeV} \lesssim M_{E_1}\ {\rm and}\  15\ {\rm GeV}\ \lesssim M_{E_1}-M_X.
\label{eq:lep-bound}
\end{align}

\begin{figure}[tbc]
\begin{center}
\includegraphics[scale=0.7]{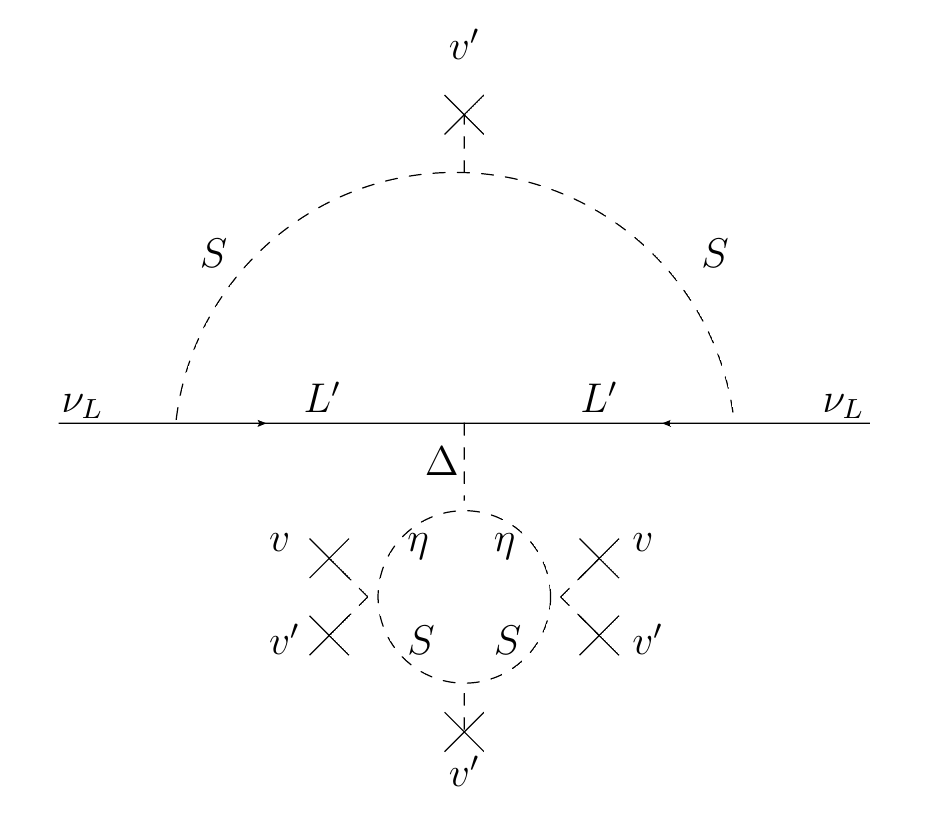}
\caption{Feynman diagram for the neutrino mass.}
\label{neut-mass}
\end{center}
\end{figure}

\subsection{Neutrino mass matrix}
The neutrino mass matrix can be generated at two-loop level  as depicted in Fig.~\ref{neut-mass}, and its form is given by
\begin{align}
({\cal M}_\nu)_{ab}=
-\frac{1}{2(4\pi)^2}
&\sum_{k}^{1,2}(y_L)_{a,k}(y_L)_{b,k}\sum_{j}^{1,2}
\left[
(V^T_{I,2j})^2
\left(\frac{M_{\psi_{1,k}}X_{1,j} }{X_{1,j}-1}\ln X_{1,j}-\frac{M_{\psi_{2,k}}X_{2,j} }{X_{2,j}-1}\ln X_{2,j}\right)\right.\nn\\
&\left.
-(V^T_{R,2j})^2
\left(\frac{M_{\psi_{1,k}}Y_{1,j} }{Y_{1,j}-1}\ln Y_{1,j}-\frac{M_{\psi_{2,k}}Y_{2,j} }{Y_{2,j}-1}\ln Y_{2,j}\right)
\right],   \label{eq:neut-theory}
\end{align}
where $X_{i,j}\equiv \left(\frac{M_{\psi_{i,k}}}{m_{A_j}}\right)^2$  and $X_{i,j}\equiv \left(\frac{M_{\psi_{i,k}}}{m_{H_j}}\right)^2$ .
Remind here that two flavor of $E_k$(k=1-2) is introduced to obtain the current neutrino oscillation data.
 $(\mathcal{M}_\nu)_{ab}$ can be generally diagonalized by the Maki-Nakagawa-Sakata mixing matrix $V_{\rm MNS}$ (MNS) as
\begin{align}
(\mathcal{M}_\nu)_{ab} &=(V_{\rm MNS} D_\nu V_{\rm MNS}^T)_{ab},\quad D_\nu\equiv (m_{\nu_1},m_{\nu_2},m_{\nu_3}),
\\
V_{\rm MNS}&=
\left[\begin{array}{ccc} {c_{13}}c_{12} &c_{13}s_{12} & s_{13} e^{-i\delta}\\
 -c_{23}s_{12}-s_{23}s_{13}c_{12}e^{i\delta} & c_{23}c_{12}-s_{23}s_{13}s_{12}e^{i\delta} & s_{23}c_{13}\\
  s_{23}s_{12}-c_{23}s_{13}c_{12}e^{i\delta} & -s_{23}c_{12}-c_{23}s_{13}s_{12}e^{i\delta} & c_{23}c_{13}\\
  \end{array}
\right],
\end{align}
where we neglect the Majorana phase as well as Dirac phase $\delta$ in the numerical analysis for simplicity.
The following neutrino oscillation data at 95\% confidence level~\cite{pdf} is given as
\begin{eqnarray}
&& 0.2911 \leq s_{12}^2 \leq 0.3161, \; 
 0.5262 \leq s_{23}^2 \leq 0.5485, \;
 0.0223 \leq s_{13}^2 \leq 0.0246,  
  \\
&& 
  \ |m_{\nu_3}^2- m_{\nu_2}^2| =(2.44\pm0.06) \times10^{-3} \ {\rm eV}^2,  \; 
  \ m_{\nu_2}^2- m_{\nu_1}^2 =(7.53\pm0.18) \times10^{-5} \ {\rm eV}^2, \nn
  \label{eq:neut-exp}
  \end{eqnarray}
where we assume one of three neutrino masses is zero with normal ordering in our analysis below.

\subsection{Muon anomalous magnetic moment and Lepton flavor violations}
The muon anomalous magnetic moment (muon $g-2$) has been 
measured at Brookhaven National Laboratory. 
The current average of the experimental results is given by~\cite{bennett}
\begin{align}
a^{\rm exp}_{\mu}=11 659 208.0(6.3)\times 10^{-10}. \notag
\end{align}
It has been well known that there is a discrepancy between the
experimental data and the prediction in the SM. 
The difference $\Delta a_{\mu}\equiv a^{\rm exp}_{\mu}-a^{\rm SM}_{\mu}$
was calculated in Ref.~\cite{discrepancy1} as 
\begin{align}
\Delta a_{\mu}=(29.0 \pm 9.0)\times 10^{-10}, \label{dev1}
\end{align}
and it was also derived in Ref.~\cite{discrepancy2} as
\begin{align}
\Delta a_{\mu}=(33.5 \pm 8.2)\times 10^{-10}. \label{dev2}
\end{align}
The above results given in Eqs. (\ref{dev1}) and (\ref{dev2}) correspond
to $3.2\sigma$ and $4.1\sigma$ deviations, respectively. 

In our model, we have new contributions to $\Delta a_\mu$ coming from $y_L$ and $y_S$ terms. 
These contributions are calculated as
\begin{align}
\Delta a_\mu &\approx
\frac{(y_L)_{21}^2
m^2_\mu}{4 (4\pi)^2}
\sum_{a}^{1-2}
\left[  (V^T_R)^2_{2a} G(H_a,E_1)+  (V^T_I)^2_{2a} G(A_a,E_1) \right]\nn\\
&
+
\frac{(y_S)_{22}^2
m^2_\mu}{4 (4\pi)^2}
\sum_{a}^{1-2}
\left[  (V^T_R)^2_{2a} G(H_a,E_2)+  (V^T_I)^2_{2a} G(A_a,E_2) \right]
\label{eq:muon-g-2}
\end{align}
where
\begin{align}
G(f_1,f_2)&\approx
\left.
 \int_0^1 dx \int_0^{1-x}dy
\frac{x(y+z)}{x m^2_{f_1}+(y+z)m^2_{f_2}}\right|_{z=1-x-y}\nn\\
&=\frac{2 m_{f_1}^6 + 3 m_{f_1}^4 m_{f_2}^2 - 6 m_{f_1}^2 m_{f_2}^4 + m_{f_2}^6 + 6 m_{f_1}^4 m_{f_2}^2\ln\left(\frac{m_{f_2}^2}{m_{f_1}^2}\right) }{6(m_{f_1}^2-m_{f_2}^2)^4}.
\end{align}

\begin{table}[t]
\begin{tabular}{c|c|c} \hline
Process & $(i,j)$ & Experimental bounds ($90\%$ CL) \\ \hline
$\mu^{-} \to e^{-} \gamma$ & $(2,1)$ &
	$\text{Br}(\mu \to e\gamma) < 5.7 \times 10^{-13}$  \\
$\tau^{-} \to e^{-} \gamma$ & $(3,1)$ &
	$\text{Br}(\tau \to e\gamma) < 3.3 \times 10^{-8}$ \\
$\tau^{-} \to \mu^{-} \gamma$ & $(3,2)$ &
	$\text{Br}(\tau \to \mu\gamma) < 4.4 \times 10^{-8}$  \\ \hline
\end{tabular}
\caption{Summary of $\ell_i \to \ell_j \gamma$ process and the lower bound of experimental data~\cite{Adam:2013mnn}.}
\label{tab:Cif}
\end{table}

Our relevant lepton flavor violation process ($\ell_i\to\ell_j\gamma$) comes from the same terms of anomalous magnetic moment at the one-loop level in principle. Each  of flavor dependent process has to satisfy the current upper bound, as can be seen in Table~\ref{tab:Cif}.
However the contribution from $y_S$ can be always negligible assuming the diagonal $y_S$. This is because this term does not contribute to the neutrino masses. Hence we consider the contribution from $y_L$ only.
Then the branching form is given as 
\begin{align}
\text{Br}(\ell_i\to\ell_j\gamma)&\approx
\sum_{k}^{1-2}\frac{6 |(y_L)_{jk} (y_L)_{ik} |^2}{(16\pi^2{\rm G_F})^2}
\left|(V_R^T)^2_{2,a}g(H_a,E_k)+(V_I^T)^2_{2,a}g(A_a,E_k)\right|^2,
\label{eq:g-2}
\end{align}
where 
\begin{align}
g(f_1,f_2)&\approx 
\left.
\int_0^1 dx \int_0^{1-x}dy
\frac{xy}{x m^2_{f_1}+(y+z)m^2_{f_2}}\right|_{z=1-x-y},
\end{align}
and ${\rm G_F}$ is Fermi constant.

\subsection{Dark matter}
First of all, we discuss the direct detection searches reported by  the experiment of LUX~\cite{Akerib:2013tjd}.
As mentioned before, the inelastic scattering process through $Z$ boson is always evaded to retain the mass difference with ${\cal O}$(100) keV, which is generated via $y'_Lv_\Delta$ in our model.
We have Higgs portal process and our scattering cross section with nucleon is given by
\begin{align}
\sigma_N\approx 0.082\frac{m^4_N |y'_L|^2}{\pi}\left[\sum_{i}^{1-3}\frac{(O_R^T)_{1i}(O^T_R)_{2i} }{m^2_{h_i}}\right]^2,
\end{align}
where $h_2\equiv h_{\rm SM}$ and the mass of neutron, which is symbolized by $m_N$, is around 939 GeV.
LUX suggests that $\sigma_N$ should be less than ${\cal O}(10^{-45})$ cm$^2$ at ${\cal O}$(10) GeV mass range of DM.

{\it Relic density of DM}: 
Before the serious analysis to the relic density, we fix a situation on DM(which is denoted by $\psi_1\equiv X$) as follows.
We focus on the region of $50\ {\rm GeV} \lesssim M_X\lesssim 80\ {\rm GeV}$ to realize a fine perspective(or simple) analysis, and suppose a co-annihilation system on DM, in which there exist  degenerated fermions $\psi_2$ and $E_1$ that could affect to the relic density ($\Omega h^2\approx0.12$) reported by Planck~\cite{Ade:2013zuv}.
Its complete analysis and formula can be found in Ref.~\cite{Griest:1990kh} and \cite{Edsjo:1997bg}. 
Our relevant processes for the thermal averaged cross section comes from the co-annihilations and annihilation for the degenerated fields
of $\Psi_{i,1} E_1\to\ell\nu_L$ via W boson, and $E_1\bar E_1\to f\bar f$ via t-channel mediated by $H_a$ and $A_a$  that comes from $y_L$ and s-channel mediated by  Z boson. Here $f$ represents all the SM fermions that are kinematically allowed.
As subdominant modes, there exist
 $X\bar X\to\nu_L\bar\nu_L$ via t- and u-channels mediated by $H_a$ and $A_a$, $X\bar\psi_2\to f\bar f$ via s-channel mediated by Z boson, $X\bar X\to2G$ via t- and u-channels mediated by $\psi_2$, 
and $X\bar X\to f\bar f$ via s-channel mediated by three CP-even neutral bosons $h_i$ from $y'_L$.
Once the total cross section $\sigma_{ij}v_{\rm rel}$ is given in al the above processes, the effective annihilation
cross section is given by 
\begin{equation}
\sigma_{\mathrm{eff}}v_{\rm rel}=\sum_{i=1}^k\sum_{j=1}^k\frac{g_{i}g_{j}}
{g_{\mathrm{eff}}^2}(\sigma_{ij}v_{\rm rel})
\left(1+\Delta_i\right)^{3/2}\left(1+\Delta_j\right)^{3/2}
e^{-(\Delta_i+\Delta_j)x},
\label{eq:coann}
\end{equation}
where $g_{\mathrm{eff}}$ is the effective degree of freedom 
\begin{equation}
g_{\mathrm{eff}}=\sum_{i=1}^kg_{i}\left(1+\Delta_i\right)^{3/2}e^{-\Delta_ix},
\end{equation}
and $\Delta_i\equiv (m_i-m_1)/m_1$ is the mass difference between DM
and the other degenerate particles, $g_i=2$ is the degree of freedom
for each Majorana particle $\chi_i$, $x=m_1/T$ and $\sigma_{ij}v$ is
(co-)annihilation cross section between $i$ and $j$.
Then the formula of  relic density is found as
\begin{align}
\Omega h^2\approx \frac{1.07\times10^9}
{g^{1/2}_* M_{\rm pl}[{\rm GeV}] \int_{x_f}^\infty \left(\frac{a_{\rm eff}}{x^2}+6\frac{b_{\rm eff}}{x^3} \right)},
\end{align}
where $g_*\approx 100$ is the total number of effective relativistic degrees of freedom at the time of freeze-out,
$M_{\rm pl}=1.22\times 10^{19}[{\rm GeV}] $ is Planck mass, $x_f\approx25$, and $ a_{\rm eff}$ and $ a_{\rm eff}$ are derived by
expanding $\sigma_{\mathrm{eff}}v_{\rm rel}$  in terms of $v_{\rm rel}$ as
\begin{align}
\sigma_{\mathrm{eff}}v_{\rm rel}\approx a_{\rm eff}+ b_{\rm rel} v^2_{\rm eff}.
\end{align}
Notice here that the mass of DM is assumed to be less than the mass of $W^\pm$ boson($\approx81$ GeV),
because the cross section is too large to satisfy the relic density once the $W^\pm$-boson final modes are open.

\section{Numerical results}
Now that all the formulae have been provided, we have a numerical analysis.
First of all, we fix the following parameters in the scalar sector:
\begin{align}
& v_\Delta \approx 0.0007686\ [{\rm GeV}],\ v' \approx 319.7\ [{\rm GeV}],\
 \mu_S \approx 278.0\ [{\rm GeV}],\ \mu_\eta \approx 715.2\ [{\rm GeV}],\ m_\eta \approx 371.8\ [{\rm GeV}],\nn\\
& m_S \approx 341.6\ [{\rm GeV}],\ 
\lambda_\Phi \approx 0.1300,\
\lambda_{\Delta_1} \approx 0.7226,\ \lambda_{\Delta_2} \approx 0.3587,\ 
\lambda_\eta \approx 0.9717,\
 \lambda_\varphi \approx 0.6274,\nn\\
 & \lambda_S \approx 0.9333,\
  \lambda_{\Phi\Delta} \approx 0.8641,\
\lambda'_{\Phi\Delta} = 0.7313,\
\lambda_{\eta\Delta}
\approx 0.9876,\
\lambda_{\eta\Delta}'\approx
0.7700,\
 \lambda_{\varphi\Delta}\approx
0.2470,\nn\\ 
&\lambda_{S\Delta}\approx
0.7253,\
\lambda_{\Phi\eta}\approx
0.6639,\
\lambda_{\Phi\eta}'\approx
0.9232,\
\lambda_{\Phi\varphi}\approx
6.912\times 10^{-6},\ 
\lambda_{S\Phi}\approx
0.7578,\nn\\
&\lambda_{\eta\varphi}\approx
 0.3144,\
\lambda_{S\eta}\approx
0.9915,\ 
\lambda_{S\varphi}\approx
0.6321,\
\lambda_{0}\approx
0.6693,\
\mu_{\rm eff} \approx 0.002806\ [{\rm GeV}],
\end{align}
where these above values satisfy the inert conditions in Eq.~(\ref{eq:cond-inert}).
Then we obtain the physical values as follows:
\begin{align}
&m_{h_{\rm SM}}\approx125.5\ [{\rm GeV}],\ m_{h_{3}}\approx368.1\ [{\rm GeV}],\ m_{h_1}\approx395.3\ [{\rm GeV}],\nn\\ 
&m_{H_1}\approx442.2\ [{\rm GeV}],\ m_{H_{2}}\approx551.7\ [{\rm GeV}],\ m_{A_1}\approx204.9\ [{\rm GeV}],\ m_{A_2}\approx454.6\ [{\rm GeV}],\nn\\ 
& m_a\approx395.3\ [{\rm GeV}],\  m_C\approx422.3\ [{\rm GeV}],\  m_{\eta^\pm}\approx417.6\ [{\rm GeV}],\  m_{\Delta^{\pm\pm}}\approx447.8\ [{\rm GeV}],\nn\\
&
O_R\approx
\left[\begin{array}{ccc}
-1 & 7.4\times 10^{-6} & 2.1\times 10^{-7}  \\
-7.4\times 10^{-6} & -1  & 4.8\times 10^{-6} \\
 2.1\times 10^{-7} & 4.8\times 10^{-6} &1   \\
\end{array}\right],\
O_I=
\left[\begin{array}{cc}
6.2\times 10^{-6}   & 1 \\
-1 &6.2\times 10^{-6}   \\
\end{array}\right],\nn\\
&
V_R\approx
\left[\begin{array}{cc}
-0.97 & 0.25   \\
 0.25 & 0.97  \\
\end{array}\right],\
V_I\approx
\left[\begin{array}{cc}
-0.16 & 0.99   \\
 0.99 & 0.16  \\
\end{array}\right],\
V_C\approx
\left[\begin{array}{cc}
4.4\times 10^{-6}  & 1   \\
 -1 & 4.4\times 10^{-6}   \\
\end{array}\right].
\end{align}
We search the other physical values with these values, where we take 
\begin{align}
& 0.9\le y'_L\le \sqrt2,\ 
10^{-6}\le (y_L)_{2,3}\le 10^{-5},
\ 50\ {\rm GeV} \le M_X \le 80\ {\rm GeV},\nn\\ 
& M_{E_1} \le M_X+45\ {\rm GeV}, \
 300 \ {\rm GeV} \le [M_{\psi_{1,2}},M_{\psi_{2,k}}]\le 500\ {\rm GeV} ,
\end{align}
in our numerical parameter spaces
to reproduce neutrino oscillation data, LFVs, and relic density of DM by using these above values.
Notice here that we solve the other $(y_L)_{i\neq2,j\neq3}$  by comparing the theoretical form and the central experimental value in Eqs.~(\ref{eq:neut-theory})-(\ref{eq:neut-exp}), and $M_{E_2}$ is assumed to be large enough to decouple these phenomenologies .
Here we show four benchmark points (BPs) as shown in Table~\ref{bench-mark}, in which we obtain ${\rm Br}(\tau\to e\gamma)$ and ${\rm Br}(\tau\to \mu\gamma)$ for all the BPs are respectively ${\cal O}(10^{-14}\sim10^{-13})$ and ${\cal O}(10^{-12}\sim10^{-11})$ that are completely safe at the current bounds. On the other hand ${\rm Br}(\mu\to e\gamma)$ is rather close to the current upper bound.
However the BP$_1$ and  BP$_2$ does not satisfy the LEP bound in Eq. (\ref{eq:lep-bound}), since the DM mass is less than 100 GeV while the mass difference between $M_X$ and $M_{E_1}$ is greater than 15 GeV. As a result, only the BP$_3$ and BP$_4$ are complete solutions to satisfy all the data that we discuss.

As for the direct detection searches, our elastic scattering cross section form reduces to $1.1\times 10^{-53} (y'_L)^2$ [cm$^2$] at the range of ${\cal O}$(10) GeV DM mass, which is always  below the current upper bound of the LUX experiment.

As for the anomalous magnetic moment, $\Delta a_\mu={\cal O}(10^{-14})$ at most  is obtained by the contribution from $y_L$ in the first term of Eq.~(\ref{eq:g-2}), which is much below the sizable value in Eq.~(\ref{dev1}) or  Eq.~(\ref{dev2}) . It originates from the fact that $y_L$ cannot enlarge due to the the constraints of neutrino oscillation data and LFV of $\mu \to e\gamma$.
On the other hand the contribution from $y_S$ in the second term of Eq.~(\ref{eq:g-2}) reaches the lower experimental value in the perturbative limit(i.e., $y_S=4\pi$), since this term can be independent of such kind of constraints.
In this limit we obtain, for example,  $\Delta a_\mu\approx2\times10^{-9}$ with $M_{E_2}=$600 GeV.

\begin{center}
\begin{table}[t]
\begin{tabular}{c||c|c|c|c|c|c|c||c|c|c}\hline\hline  
& ~$|(y_L)_{1,1}|$~ & ~$|(y_L)_{1,2}|$~ &~$|(y_L)_{1,3}|$~&~$|(y_L)_{2,1}|$~&  ~$|(y_L)_{2,2}|$~ & ~$|(y_L)_{2,3}|$~ & ~$\frac{M_{X}}{\rm GeV}$ 
&~$\frac{M_{E_1}}{\rm GeV}$  & ~$\Omega h^2$ & ~${\rm Br}(\mu\to e\gamma)$ \\\hline 
${\rm BP}_1$ 
& $0.0025$  & $0.031$ & $0.036$ & $1.6\times10^{-4}$ & $2.7\times10^{-4}$ &$4.5\times10^{-6}$
& ${62.27}$&${79.55}$ &$0.119$
& $8.1\times 10^{-14}$ \\\hline 
${\rm BP}_2$  
& $0.0029$  & $0.028$ & $0.032$ & $2.5\times10^{-5}$ & $3.9\times10^{-5}$ &$4.3\times10^{-6}$
& ${59.65}$&${76.49}$ &$0.120$
& $9.6\times 10^{-14}$ \\\hline
${\rm BP}_3$  
& $0.0023$  & $0.029$ & $0.034$ & $5.4\times10^{-5}$ & $9.5\times10^{-5}$ &$4.5\times10^{-6}$
& ${78.67}$&${103.6}$ &$0.112$
& $7.3\times 10^{-14}$ \\\hline
${\rm BP}_4$  
& $0.0014$  & $0.0216$ & $0.019$ & $2.9\times10^{-5}$ & $5.1\times10^{-5}$ &$4.7\times10^{-6}$
& ${79.7}$&${100.5}$ &$0.125$
& $7.8\times 10^{-15}$ \\\hline 
\end{tabular}
\caption{Four bench mark points(BPs), where ${\rm Br}(\tau\to e\gamma)$ and ${\rm Br}(\tau\to \mu\gamma)$ for all the BPs are respectively ${\cal O}(10^{-14}\sim10^{-13})$ and ${\cal O}(10^{-12}\sim10^{-11})$ that are completely safe.
}
\label{bench-mark}
\end{table}
\end{center}

\section{Conclusions}
We have studied a two-loop induced radiative neutrino model, in which we have shown some allowed bench mark points to satisfy
the observed neutrino masses, LFVs, and the relic density of DM in the co-annihilation system satisfying the current upper bound  on the spin independent scattering with nucleon as well as LEP.
We have also shown a new source ($y_S$) to marginally obtain the sizable value of the anomalous magnetic moment  in the perturbative limit.

\section*{Acknowledgments}
\vspace{0.5cm}
Authors thank to Dr. Kei Yagyu for fruitful discussions.
H.O. expresses his sincere gratitude toward all the KIAS members, Korean cordial persons, foods, culture, weather, and all the other things.
This work was supported by the Korea Neutrino Research Center which is established by the National Research Foundation of Korea(NRF) grant funded by the Korea government(MSIP) (No. 2009-0083526) (Y.O.).

\end{document}